\documentclass[sigconf,nonacm]{acmart}

\usepackage{tikz}
\usepackage{xspace}
\usepackage{paralist} 

\usepackage{todonotes}
\usepackage{marginnote}
\let\marginpar\marginnote

\newcommand{\parhead}[1]{\textbf{#1.}}
\newcommand{\numAttacks}{13\xspace}
\newcommand{\numPapers}{7\xspace}

\newcounter{researchquestion}
\newcommand{\defrq}[1]{\refstepcounter{researchquestion}\label{#1}RQ\theresearchquestion}
\makeatletter
\newcommand{\researchquestionautorefname}{RQ\@gobble}
\makeatother

\newcommand*{\circled}[1]{%
  \tikz[baseline=(char.base)]{\node[shape=circle,draw,minimum size=3.5mm,inner sep=0pt] (char) {#1};}%
}

\newcommand{\countermeasure}{\textsc{PreFence}\xspace}

\usepackage{array}
\usepackage{ragged2e}
\newcolumntype{L}[1]{>{\RaggedRight\hspace{0pt}}p{#1}}
\newcolumntype{R}[1]{>{\RaggedLeft\hspace{0pt}}p{#1}}
\newcolumntype{C}[1]{>{\Centering\hspace{0pt}}p{#1}}

\begin{document}

\title{A Scheduling-Aware Defense Against Prefetching-Based Side-Channel Attacks}

\author{Till Schlüter}
\affiliation{%
  \institution{CISPA Helmholtz Center for Information Security}
  \city{Saarbrücken}
  \country{Germany}}
\email{till.schlueter@cispa.de}

\author{Nils Ole Tippenhauer}
\affiliation{%
  \institution{CISPA Helmholtz Center for Information Security}
  \city{Saarbrücken}
  \country{Germany}}
\email{tippenhauer@cispa.de}

\begin{abstract}
  Modern computer processors use microarchitectural optimization mechanisms to improve performance.
  As a downside, such optimizations are prone to introducing side-channel vulnerabilities.
  Speculative loading of memory, called \emph{prefetching}, is common in real-world CPUs and may cause such side-chan\-nel vulnerabilities: Prior work has shown that it can be exploited to bypass process isolation and leak secrets, such as keys used in RSA, AES, and ECDH implementations. However, to this date, no effective and efficient countermeasure has been presented that secures software on systems with affected prefetchers.

  In this work, we answer the question: \emph{How can a process defend against prefetch-based side channels?}
  We first systematize prefetching-based side-channel vulnerabilities presented in academic literature so far.
  Next, we design and implement \countermeasure, a scheduling-aware defense against these side channels that allows processes to disable the prefetcher temporarily during security-critical operations.
  We implement our countermeasure for an x86\_64 and an ARM processor; it can be adapted to any platform that allows to disable the prefetcher. 
  We evaluate our defense and find that our solution reliably stops prefetch leakage. 
  Our countermeasure causes negligible performance impact while no security-relevant code is executed, and its worst case performance is comparable to completely turning off the prefetcher. The expected average performance impact depends on the security-relevant code in the application and can be negligible as we demonstrate with a simple web server application.

  We expect our countermeasure could widely be integrated in commodity OS, and even be extended to signal generally security-relevant code to the kernel to allow coordinated application of countermeasures.
\end{abstract}


\maketitle


\section{Introduction}
\label{sec:introduction}

Prefetching is an optimization mechanism of modern CPUs that aims to bring chunks of memory into the cache before they are actually loaded by application code. By bringing those chunks closer to the CPU in the memory hierarchy, applications benefit from lower memory latency.
There are two kinds of prefetching: software prefetching and hardware prefetching.
Software prefetching relies on explicit hints issued by application software to indicate which memory locations are likely to be accessed in the near future.
In contrast, hardware prefetching is an automatic and fully transparent mechanism that analyzes memory accesses at runtime, tries to detect regular or recurring patterns and tries to predict memory locations that are likely to be accessed soon.

Most hardware prefetchers keep an internal state that controls the prefetching process. After the prefetcher has observed secret-dependent memory accesses, its state may correlate with those secrets. Prior work has shown that the prefetcher's subsequent behavior can then be exploited as a side channel to compromise Diffie-Hellman keys~\cite{shin-prefetch,chen:2024:gofetch}, RSA private keys~\cite{afterimage,chen:2024:gofetch}, or AES symmetric keys~\cite{xiao:2023,fetchbench}.
In addition, covert channels based on hardware prefetching have been presented~\cite{rohan:2020,cronin:2019,afterimage,fetchbench}, in some cases bypassing process isolation guarantees.
A covert channel is a hidden communication channel between a sender and a receiver, both controlled by an attacker. In contrast, in a side-channel attack, only the receiving end is controlled by the attacker, while a victim process (involuntarily) acts as the sender.
No defense has been presented to date that protects against prefetcher-based attacks effectively and and efficiently.
For example, while most platforms allow to disable the prefetcher completely, this will have significant performance impact on non-security relevant parts of the applications, and all parallel applications that share the prefetcher.

In this paper, we propose \countermeasure, our novel countermeasure that allows an application to defend itself against the perils of hardware prefetching with minimal overhead. 
More precisely, our solution comes with negligible performance overhead on non-security-critical workloads.
For security-critical workloads, the worst-case performance impact is comparable to disabling the prefetcher completely, but can also be negligible overall, depending on the protected workload and the way our countermeasure is applied.

We systematically analyze existing side-channel attacks that exploit hardware prefetching for their differences and similarities and we identify suitable entry points for defenses. Based on these insights, we design, implement and evaluate \countermeasure: a mechanism that allows applications to disable the prefetcher temporarily during security-critical operations. We also address the challenges arising from process scheduling and related to multi-core processing and Simultaneous Multithreading (SMT).
As a software-based mitigation, \countermeasure leverages the widespread support of processors to disable the prefetcher and does not require further hardware adaptations.
We focus on defending processes against falling victim to side-channel attacks based on hardware prefetching, as those attacks directly expose secrets from the victim's context; we exclude covert channels, as those can merely be used to transfer information that is already accessible to the attacker.

\parhead{Contributions}
We summarize our contributions as follows:
\begin{compactitem}
  \item We systematize existing prefetch-based side channel attacks and identify their similarities and differences. We identify 5 main stages and map each attack's flow to those stages, demonstrating that there are core components required by all attacks.
  \item We design, implement and evaluate \countermeasure, our approach to mitigate prefetch-based side channels in the scheduler. We demonstrate the performance impact is negligible, and that it prevents a prior work attack.
  \item We review software-based mitigations proposed in prior work and argue why they are either incomplete (e.g., do not consider SMT) or too costly (e.g., permanently disable the prefetcher, rewriting code as constant time). \countermeasure fills this gap.
\end{compactitem}

We provide an open-source implementation of \countermeasure at \url{https://github.com/scy-phy/PreFence/tree/preprint}.

\section{Background}
\label{sec:background}

\subsection{Caches and Prefetching}
\parhead{Caches}
Modern processors aim to reduce the effective latency of memory accesses by maintaining \emph{caches}. A cache is a fast and small temporary storage that stores frequently or recently used chunks of memory.
These chunks are called cache lines and have a fixed size.
When a program loads data from a memory address, the processor first checks whether the data is present in a cache (\emph{cache hit}) or not (\emph{cache miss}). In case of a hit, the load is significantly faster. Otherwise, the data needs to be fetched from DRAM, which takes more time.

\parhead{Prefetching}
Apart from chunks of memory that have been used in the past, modern processors may also bring chunks of memory into the cache that are likely to be accessed in the near future. To this end, a hardware unit of the processor, the prefetcher, observes memory accesses at runtime and predicts addresses that are likely to be accessed next. Those predictions are often generated by undocumented prediction mechanisms~\cite{fetchbench}.
While most prefetchers only analyze addresses to generate predictions, more powerful \emph{data memory-dependent prefetchers (DMPs)} also take the memory contents into account~\cite{augury,chen:2024:gofetch}.
Prefetching is a completely transparent mechanism from the application's point of view.
To make useful predictions, most prefetchers keep an internal state that reflects recent memory activity. They are often implemented as a shared resource between processes running on the same physical processor core.
These properties make prefetchers susceptible to side-channel vulnerabilities, as we discuss in detail in Section~\ref{sec:systematization}.

\subsection{Simultaneous Multithreading (SMT)}
Traditionally, every processor core executes exactly one program thread at a time. On a system that supports multithreading, multiple threads take turns in using the core.
To switch from one thread to another, the state of the current thread needs to be stored in memory and the state of the next thread needs to be restored to the processor registers. This procedure is known as \emph{context switching} and handled by the scheduler, a component of the operating system~\cite{tanenbaum:2015}.

Simultaneous Multithreading (SMT)~\cite{tullsen:1995:smt} is a concept that aims to better utilize the resources of a processor core.
The idea behind SMT is to schedule multiple threads on a single processor core \emph{at the same time}, based on the insight that a single thread is often not able to utilize all the processing units that a processor core has available.

SMT has been adopted by major processor vendors such as Intel (branded ``HyperThreading'')~\cite{marr:2002:hyperthreading} and AMD~\cite{clark:2016:amd-smt}. These practical implementations expose one physical processor core as multiple (often two) independent logical cores to the operating system. We refer to logical cores that are backed by the same physical core as \emph{sibling cores}. The operating system schedules threads on logical processors in the same way as it would on a traditional system~\cite{marr:2002:hyperthreading}.

On a non-SMT system, a thread has exclusive access to the resources of a processor core while it is scheduled.
In contrast, on an SMT-enabled system, the instructions issued by parallel threads scheduled on sibling cores share processor resources at the same time, potentially also the prefetcher.
We emphasize that, as a result of SMT, instructions issued by multiple processes can be executed on the same physical core without requiring a context switch.

\section{Defending Against Prefetching-\\Based Side-Channel Attacks}
\label{sec:three}

\subsection{System and Attacker Model}
\label{sec:model}
We assume a system with a processor that performs prefetching. We further assume that the defender and attacker know the type of the deployed prefetcher, as well as its security-relevant characteristics (e.g. obtained by the attacker with a copy of the target hardware and a suitable testbench~\cite{fetchbench}).
The defender is able to modify the software running on the CPU, including the operating system kernel.
The hardware provides an interface to control (i.e., enable or disable) the prefetcher from the kernel.
The attacker is able to execute arbitrary code in userspace. In Section~\ref{sec:discussion:scope}, we extend the attacker model to attackers at higher privilege levels.

\subsection{Research Questions and Challenges}
\label{sec:rq}

In this work, we answer the following research questions:
\begin{compactitem}
  \item[\textbf{\defrq{rq:vulns}:}] What kind of side-channel vulnerabilities in prefetchers have been exploited in prior work? Is there a core set of vulnerabilities that are critical for all known attacks?
  \item[\textbf{\defrq{rq:countermeasure}:}] Is there a software-only countermeasure to mitigate all known prefetching-based side-channel vulnerabilities effectively and efficiently?
  \item[\textbf{\defrq{rq:mitigations}:}] 
  How can prior work on countermeasures be systematized, and which attacks can be expected to be prevented by those defenses?
\end{compactitem}

\parhead{Challenges}
To answer these research questions, we need to overcome the following challenges:
\begin{compactenum}
  \item Prefetcher side channels have been exploited in different settings in prior work. We need to work out similarities and differences between those approaches to be able to identify common patterns.

  \item Any countermeasure will cause a performance impact, which will need to be quantified and minimized.

  \item Countermeasures require trustworthy arguments on why they can be expected to prevent current and future attacks. So far, such arguments have not been provided in prior work and across different attacks. 
\end{compactenum}

\parhead{Proposed Approach}
To overcome these challenges, we pursue the following approach.
First, we systematize known prefetching-based side-channel attacks and their exploited vulnerabilities. We identify a minimal set of vulnerabilities that are required for any attack to work.
Second, we design and implement a solution to prevent exploitation of this minimal set of vulnerabilities, leading to a countermeasure effective against all prefetcher side-channel attacks from userspace. We then evaluate the implemented solution on real-world hardware.
Finally, we review software-based mitigations from prior work and discuss their efficacy and efficiency.

\section{Systematization of Attacks}
\label{sec:systematization}

To protect against prefetching-based side channels, we first need to understand the attack vectors in detail.
To this end, we systematize all attacks exploiting hardware-based data prefetchers that we could find in academic literature (\numAttacks attacks across \numPapers papers). We list them in Table~\ref{tab:attacks}.
Inspired by prior works on mitigating other microarchitectural side channels~\cite{canella:2019,specfuscator,msft-speculative}, we break down prefetcher-based attacks into stages. We further define the relevant scopes that those attacks operate in and report them in Table~\ref{tab:attacks}.
Finally, we visualize our systematization by plotting the attack sequences in Figure~\ref{fig:attacks-flow}, deduce similarities and differences, and expose where software-based mitigations can effectively be applied.

\subsection{Stages of Prefetching Side Channels}

\begin{figure}
  \centering
  \includegraphics[width=0.8\linewidth]{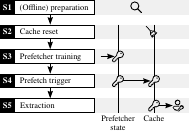}
  \caption{Stages of prefetching-based side channels}
  \label{fig:stages}
\end{figure}

We observe that prefetching side channels can be split into the following five stages, as illustrated in Figure~\ref{fig:stages}:
\begin{compactitem}
  \item[\textbf{S1:}] \textbf{(Offline) Preparation.} For some attacks, an attacker has to take preliminary steps before the actual attack begins, such as reverse engineering or setting up data structures.
  \item[\textbf{S2:}] \textbf{Cache Reset.} To start from a clean cache state, some attackers  perform actions that reset the initial cache state.
  \item[\textbf{S3:}] \textbf{Prefetcher Training.} Most prefetchers keep an internal state that determines their behavior. The prefetcher observes memory accesses at runtime and tries to identify patterns in the addresses or data being accessed. If a pattern is detected or a previously detected pattern is continued or interrupted, the prefetcher's internal state may be altered to change the prefetcher's future behavior.
  We refer to this state change as \emph{prefetcher training} throughout this paper. From the attacker's perspective, this step can be seen as \emph{encoding information into the prefetcher's state}. In the context of an attack, this information is secret-dependent.
  \item[\textbf{S4:}] \textbf{Prefetch Trigger.}
  Upon a trigger event, such as another memory access that matches certain criteria, a prefetcher may bring additional memory lines into the cache. Those memory lines are selected based on the prefetcher's internal state.
  We refer to this process as \emph{prefetch trigger} throughout this paper. From the attacker's perspective, this step can be seen as \emph{extracting information from the prefetcher's state into the cache}.
  \item[\textbf{S5:}] \textbf{Extraction.} The cache state is inspected to extract information about the prefetcher's internal state. This step can be seen as \emph{extracting information from the cache into the attacker's context}.
\end{compactitem}
Attacks may skip some of these stages (see Figure~\ref{fig:attacks-flow}).

\subsection{Scopes}
Prefetch attacks often operate across privilege domains. In this respect, we classify attacks based on the following scopes:

\begin{compactitem}
  \item[\textbf{SP:}] \textbf{Same-process.} Leaking within the same process.
  \item[\textbf{CT:}] \textbf{Cross-thread.} Leaking from one thread of a userspace process to another.
  \item[\textbf{CP:}] \textbf{Cross-process.} Leaking from userspace process to another.
  \item[\textbf{KU:}] \textbf{Kernel to user.} Leaking from kernel to user\-space.
  \item[\textbf{TO:}] \textbf{TEE to OS.} Leaking from a trusted execution environment (TEE), such as Intel SGX or ARM TrustZone, to the (untrusted) operating system.
\end{compactitem}

If victim and attacker do not share the same context (e.g., they are different userspace processes), the attacker needs to make sure that the prefetcher keeps its state across the context switch. Especially when leaking between two userspace threads or processes, the attacker faces the problem that the scheduler manages the process runtime, making it non-trivial to interrupt the victim process at a specific point in time (when the prefetcher's state is secret-dependent) and schedule the attacker process (to extract the state). Some attacks assume shared memory, either data memory or shared libraries, between both processes to address this issue~\cite{afterimage}, others use additional side channels for synchronization~\cite{fetchbench}.

\subsection{Prefetching Attack Systematization}
\label{sec:attacks}
\label{sec:attacks-conclusion}

\begin{table}
  \caption{Overview of prefetching-based attacks in prior work}
  \label{tab:attacks}
  \centering
  \footnotesize
  \begin{tabular}{rllll}
    \toprule
    \textbf{Attack}
      & \textbf{Prefetcher}
      & \textbf{Scope}
      & \textbf{Target}\\
    \midrule
    Shin et~al.~\cite{shin-prefetch}
      & Intel IP stride
      & CP
      & OpenSSL ECDH\\
    Augury~\cite{augury} OOB
      & Apple DMP
      & SP
      & Custom\\
    Augury~\cite{augury} SLH
      & Apple DMP
      & SP
      & Custom\\
    Augury~\cite{augury} Addr.
      & Apple DMP
      & SP
      & ---\\
    AfterImage~\cite{afterimage} Var.~1
      & Intel IP stride
      & CT/CP
      & Custom\\
    AfterImage~\cite{afterimage} Var.~2
      & Intel IP stride
      & KU
      & Custom\\
    AfterImage~\cite{afterimage} SGX
      & Intel IP stride
      & TO
      & Custom\\
    AfterImage~\cite{afterimage} RSA
      & Intel IP stride
      & CT
      & MbedTLS RSA\\
    AfterImage~\cite{afterimage} Sync
      & Intel IP stride
      & CP
      & OpenSSL RSA\\
    Xiao et~al.~\cite{xiao:2023}
      & Intel IP stride
      & SP
      & AES\\
    FetchBench~\cite{fetchbench} AES
      & ARM SMS
      & CP
      & MbedTLS AES\\
    PrefetchX~\cite{chen:2024:prefetchx}
      & Intel XPT
      & CP
      & MbedTLS RSA, GnuPG RSA\\
    GoFetch~\cite{chen:2024:gofetch}
      & Apple DMP
      & CP
      & Go RSA, OpenSSL DHKE,\\
      &&& CRYSTALS\\
    \bottomrule
  \end{tabular}
\end{table}

\begin{figure*}
  \includegraphics[width=\textwidth]{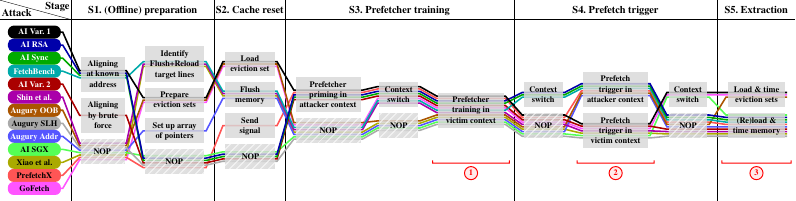}
  \caption{Overview of the sequence of activities in prefetching-based attacks. Core activities required by all highlighted in red.}
  \label{fig:attacks-flow}
\end{figure*}

We now map \numAttacks attacks from prior work to the five stages introduced above (details in Appendix~\ref{sec:attacks-detail}) and answer RQ\ref{rq:vulns}. We summarize the results in Figure~\ref{fig:attacks-flow}, showing the sequence of activities per attack. If an attack does not perform any of the alternative activities at some point in the sequence, we record a transition through a \emph{no-operation (NOP)} block.

\parhead{Finding: There Are Mandatory Stages}
Our systematization shows that prefetcher training~\circled{1}, prefetcher triggering~\circled{2}, and cache extraction~\circled{3} (highlighted in red in Figure~\ref{fig:attacks-flow}) are activities that are common to all attacks, as there are no \emph{NOP} alternatives.

\parhead{Finding: Prefetch Attacks Are Cache Attacks}
We further note that all prefetcher-based side channels are cache-based attacks: All attacks are built upon techniques to probe the cache state of certain cache lines~\circled{3}, typically akin to Flush+Re\-load~\cite{flush-reload} and Prime+Probe~\cite{osvik:2006}.

\parhead{Finding: Prefetching Is Triggered by Victim or Attacker}
We identify that many attacks rely on the victim context to trigger the prefetcher to extract information from the prefetcher's state and transfer it into the cache~\circled{2}. However, recent works have shown that this step can be moved into the attacker's context in some cases, even though this involves more complex synchronization steps~\cite{afterimage,fetchbench,chen:2024:prefetchx}.

\parhead{Finding: Victim Trains The Prefetcher}
Most importantly, we emphasize that all attacks rely on prefetcher training within the victim context~\circled{1}.
This is plausible, as the victim necessarily needs to work with the secret (e.g., perform secret-dependent memory accesses) to encode it into the prefetcher's state.
Notably, from a defender's perspective, this means that the victim is able to protect itself using mitigations applied to its own code.

We conclude that a general mitigation approach against prefetch-based side-channel attacks is to \emph{ensure that no training occurs in the victim context}.
Fortunately, code running in victim contexts is easier to control than attacker code, which is (as the name suggests) constituted by the attacker.

\section{\countermeasure: Design and Implementation}
\label{sec:countermeasure}
We now answer RQ\ref{rq:countermeasure} by presenting \countermeasure, our software-only countermeasure against prefetching-based side-channel attacks. It exploits that the victim process trains the prefetcher in all attacks.
We discuss alternative software-based defense approaches and why we consider them infeasible in Section~\ref{sec:prior-countermeasures}.

\parhead{Design Goals}\label{sec:design-goals}
The main design goals of \countermeasure are:
\begin{compactitem}
  \item[\textbf{DG1:}] It mitigates all prior prefetch-based attacks conducted from userspace
  \item[\textbf{DG2:}] It is simple to use for application developers and end users
  \item[\textbf{DG3:}] It has minimal runtime overhead
  \item[\textbf{DG4:}] It is still functional when the prefetcher is shared across physical or SMT sibling cores
\end{compactitem}
We show that \countermeasure achieves these goals in the following sections.

\parhead{Approach: Disabling the Prefetcher Temporarily}
We take the approach of disabling the prefetcher temporarily, for example while security-critical code is executed.
As we have shown in Section~\ref{sec:attacks-conclusion}, stage S3 is the critical phase in all attacks. In this stage, secrets are encoded into the prefetcher's state. To achieve DG1, we exploit (and verify in Section~\ref{sec:disabled-behavior}) that the prefetcher does not update its state while it is disabled, i.e., it cannot be trained in this state. 
By disabling the prefetcher temporarily while secrets are processed, we ensure that the prefetcher is not trained with secrets. Afterward, the prefetcher can be enabled again. This limits the performance impact of the missing prefetching mechanism to the security-critical code and thus enables DG3.

\parhead{Challenges}
While this countermeasure sounds straightforward to implement at first, some challenges become apparent on closer inspection.
First, a security-critical process that requests the prefetcher to be disabled may be interrupted by the scheduler and replaced by another process. In that case, the prefetcher must be re-enabled while the other process is running and disabled again when the security-critical process is re-scheduled. This ensures
(i) that the interrupting process can benefit from the prefetcher's performance boost, and
(ii) that the interrupting process cannot maliciously re-enable the prefetcher.
Second, a security-critical process may be migrated from one processor core to another. If the prefetcher operates on a per-core basis, it needs to be re-enabled on the original core and disabled on the destination core.
Third, if the prefetcher is shared across physical or SMT sibling cores, a malicious process that runs concurrently on another core could re-enable the prefetcher while the security-critical process is running. Consequently, the prefetcher must not be enabled if any of the processes sharing the same prefetcher requested it to be disabled.
Lastly, access to MSRs that control the prefetcher is usually not allowed from userspace.

\subsection{Design}
We conclude from the above challenges that our countermeasure needs to be tightly integrated with the operating system kernel and the scheduler. In that way, relevant registers can be accessed and it can be ensured that the prefetcher is disabled as long as a program requests it, regardless of intermittent scheduling events.

\parhead{Victim-Initiated Prefetch Control}
Our countermeasure empowers a userspace process to protect itself from prefetching-based side channel attacks by requesting the prefetcher to be disabled temporarily.
More precisely, a process signals to the operating system kernel when it enters or leaves a security-critical code section, such as an encryption function. While this requires developers to add small amounts of code, the required changes are minimal and the effort is low compared to complex re-writes required by other countermeasures (such as constant-time programming, see Section~\ref{sec:prior-countermeasures}). In this way, \countermeasure achieves DG2.
The kernel keeps track whether a process is currently in a security-critical code section and ensures that the prefetcher is disabled during this period---even if it is interrupted by the scheduler.
After disabling the prefetcher, it remains disabled until the requesting process is finished with executing its security-relevant code, or a different process without security-relevant code is scheduled into.
If the next-to-be-scheduled process also requested to have the prefetcher disabled, the prefetcher remains disabled until non-critical code is reached.
This is enforced by the scheduler, which changes the prefetcher's state based on the request of the next process to be executed.

\begin{figure}
  \includegraphics[width=\linewidth]{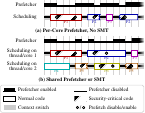}
  \caption{\countermeasure at work: The prefetcher is disabled temporarily while security-critical code is executed. On SMT-capable cores and for shared prefetchers, the scheduler considers requested states of the relevant parallel processes.}
  \label{fig:countermeasure}
\end{figure}

\parhead{The Simple Case: Per-Core Prefetcher, No SMT}
We illustrate our countermeasure by example of Figure~\ref{fig:countermeasure}~(a), starting with the simple case in the upper half: a prefetcher that is exclusive to one core on a CPU without SMT.
When a process is started, prefetching is enabled by default. As process P1 shows, the process can then request prefetching to be disabled, perform a security-critical operation, and request prefetching to be enabled again.
Process P2 illustrates the case where a process is interrupted by the scheduler during a security-critical code section. The scheduler deschedules P2 and checks whether the next process requested prefetching to be disabled or not. In this example, the next processes (the new process P3) did not request prefetching to be disabled. Consequently, the scheduler re-enables prefetching while P3 is running. Once P3 is finished, the scheduler disables the prefetcher again and switches back to P2.

\parhead{The Special Case: Shared Prefetcher or SMT}
If the processor uses SMT and sibling cores share the same prefetcher, we need to be careful about implicit context switches in order to fulfill DG4. With SMT, multiple processes can be executed on the same physical processor core simultaneously without operating-system-controlled context switches between them. In the worst case, when an attacker and a victim process are scheduled on sibling cores, the victim could request the prefetcher to be disabled, while the attacker requests it to be enabled again. Similarly, if the prefetcher is shared across cores, an attack could be run simultaneously from a different core.
For this reason, it is not enough to update the prefetcher's state on OS-controlled context switches in these cases. Instead, we keep prefetching disabled on all (logical) cores that share a prefetcher as soon as and as long as any of the scheduled processes requests it.

We provide an example in the lower half of Figure~\ref{fig:countermeasure}.
First, process P1 requests prefetching to be disabled while the parallel process P4 does not request it. To protect process P1, prefetching is disabled until P1 leaves the security-critical code section.
Next, process P5 enters a security-critical section, but is soon interrupted by the scheduler. Since no more processes request the prefetcher to be disabled after P5 is descheduled, the prefetcher is enabled again.
Next, processes P2 and P6 both run security-critical code in parallel. The prefetcher is disabled when the first process (P2) enters the security-critical section and is re-enabled when the last process (P6) leaves it.
Finally, P5 is re-scheduled. As it was interrupted in a security-critical section, the prefetcher is disabled at context switch and re-enabled once the security-critical section is completed.

\parhead{Core Migrations}
We note that our methodology also handles the case of a core migration. On a context switch, the scheduler adjusts the prefetcher's state based on the request of the next process. When a core migration occurs, the prefetcher's state on the original core is determined by the next process running on that core. On the target core, the migrating process is the next process, so it decides about the prefetcher's state.

\subsection{Implementation}
\label{sec:impl}
\parhead{\countermeasure Kernel Patch}
We implement our countermeasure as a Linux kernel patch. Our prototype is currently able to control the prefetchers of Intel x86\_64 (tested on Comet Lake) and ARM Cortex-A72 CPUs. Excluding comments, our patch adds only 91 (Intel) / 62 (ARM) lines of code to the Linux kernel code base.
We extend the \texttt{task\_struct}, the place where the scheduler keeps all information related to a process, with a boolean \texttt{prefetch\_\-disable} flag. The flag is initialized to \texttt{false} for new processes, i.e., prefetching is enabled by default and can be disabled on request.

To allow processes to control this flag from userspace, we add options to set, clear or query the flag to the \texttt{prctl} system call. When the flag is changed through the system call, the kernel updates the \texttt{task\_struct} of the calling process accordingly. In addition, the kernel changes the prefetcher's state on the respective CPU immediately by writing to the corresponding MSRs~\cite{intel-prefetcher-disclosure} before returning to userspace.
We further extend the scheduler's \texttt{context\_switch} function to update the state of the prefetcher on context switches based on the prefetch-disable flag of the next process.

To deal with prefetchers shared across physical or SMT sibling cores, we keep a global bit vector (of type \texttt{cpumask\_t}) that indicates for each CPU whether it currently runs a process with the prefetch-disable flag set. We check this bit vector before enabling the prefetcher on any core. Only if none of the cores sharing the same prefetcher currently runs a process with the prefetch-disable flag set, the prefetcher can be enabled; otherwise, it remains disabled.

\parhead{Using \countermeasure in Applications}
To make use of \countermeasure, the \texttt{prefetch\_\-disable} flag needs to be set before entering security-critical code sections and cleared afterward (through system calls).
The countermeasure can be applied at various levels.
For instance, library developers can protect security-critical code sections, for example those that process secrets. This ensures that the prefetcher is disabled only for a minimal period of time.
Application developers who use unprotected legacy libraries can resort to setting the bit themselves before performing security-critical library calls and clearing it afterward. In this case, the prefetcher is disabled for a longer but still limited period of time.

\countermeasure can even be made available to knowledgeable end users who want to protect legacy software. A wrapper program similar to \texttt{taskset} can set the \texttt{prefetch\_\-disable} flag in advance and then execute a target application. The target application inherits the flag and is thus executed with prefetching disabled. This means that the target application cannot benefit from prefetching at all. However, the impact on the overall system performance is still limited to one application, and other applications can still benefit from prefetching.

\section{Evaluation}
In this section, we evaluate our \countermeasure implementation for efficacy and efficiency.
First, as a prerequisite, we evaluate the behavior of a disabled prefetcher to verify that disabling the prefetcher has the expected effects (no further training), and thus \countermeasure is applicable. 
Next, we demonstrate the efficacy of our countermeasure. It prevents the prefetching-based side channel presented by Shin et al.~\cite{shin-prefetch} (reproduced by us).
Finally, we show that \countermeasure is also efficient. We investigate this in three scenarios:
\begin{itemize}
  \item \parhead{Scenario 1: Stock kernel}
  For an unmodified (``stock'') kernel, we measure the performance impact when the prefetcher is disabled for the whole execution time of an application (using SPEC benchmarks).
  These measurements serve as a baseline for the following experiments.
  \item \parhead{Scenario 2: Patched kernel with non-critical workload}
  For a patched kernel, we measure the performance impact (on SPEC benchmarks) when no process makes use of the possibility to disable prefetching.
  These measurements show the fixed performance overhead on a process introduced by our countermeasure.
  \item \parhead{Scenario 3: Patched kernel with critical workload}
  As an end-to-end example, we evaluate the performance of a web server application running on a patched kernel.
  Using our syscall, we disable the prefetcher during the TLS-related code execution.
  In addition to cryptographic code, each HTTP request to our server also triggers application code which can still benefit from the prefetcher (e.g., a database lookup).
  We  test different values for the complexity of the database lookup to find at which point the performance impact of disabling the prefetcher during TLS parts becomes negligible.
\end{itemize}
To investigate the overhead further, we specifically evaluate the introduced fixed overhead on every context switch and the one-off overhead of a system call whenever a \texttt{pre\-fetch\_\-dis\-able} flag is modified in isolation. These additional results can be found in Appendix~\ref{sec:appendix-eval}.

\subsection{Evaluation Environments}
For all experiments in the main body of this paper, we use the following two platforms throughout the evaluation.

\parhead{x86\_64}
Our x86\_64 platform is an Intel Core i7-10510U (Comet Lake) CPU running Alpine Linux 3.19 with kernel \texttt{6.6.14-\allowbreak{}r0-\allowbreak{}lts}. Depending on the experiment, we either use the original kernel from the Alpine repositories or our patched kernel derived from it. We use the \texttt{rdtscp} instruction to measure time.

\parhead{ARM}
Our ARM platform is a Raspberry Pi~4 using a Broadcom BCM2711 SoC with four Cortex-A72 cores. It runs Raspberry Pi OS~12 64-bit with Linux kernel \texttt{6.6.22-\allowbreak{}v8}. We either use a kernel that we compiled from the official sources~\cite{rpi-kernel-docs} without any changes or a kernel derived from it using our kernel patch. We use the cycle count register (\texttt{PMCCNTR\_EL0}) to measure time.

\subsection{Prerequisite: Disabled Prefetcher Behavior}
\label{sec:disabled-behavior}

\begin{figure*}
  \includegraphics[width=\textwidth]{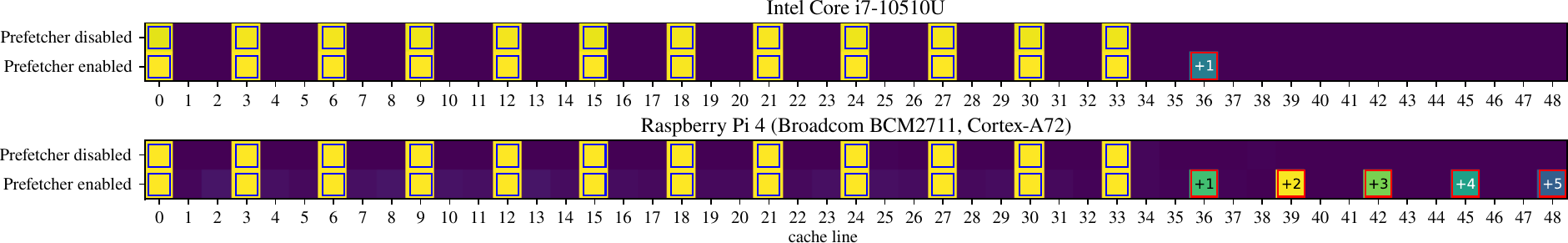}
  \caption{Prefetcher behavior when trained while disabled (compared to the behavior when trained while enabled). Blue boxes denote accesses, red boxes indicate prefetch locations. The tested prefetchers cannot be trained while it is disabled.}
  \label{fig:disabled-behavior}
\end{figure*}

\countermeasure requires that a disabled prefetcher cannot be trained, i.e., it does not update its state while it is disabled.

\parhead{Experiment}
To test the prefetcher for this behavior, we implement a corresponding testcase for stride prefetchers, the most common type of prefetchers, in the FetchBench framework~\cite{fetchbench}.
We first disable the prefetcher, access a sequence of memory locations with constant distance between them, re-enable the prefetcher, and perform one more memory access matching the pattern. If the prefetcher keeps learning while disabled, we expect it to be triggered by that last access and bring more elements into the cache. Otherwise, no prefetching effects should appear in cache. As a baseline, we repeat the same experiment with the prefetcher being enabled. In that case, we expect prefetching effects in the cache.

\parhead{Results}
We run the testcase on the Intel Core i7-10510U and BCM2711 processors. As illustrated in Figure~\ref{fig:disabled-behavior}, we find that the prefetchers cannot be trained while disabled.
We conclude that \countermeasure is applicable to these prefetcher implementations.

\subsection{Efficacy: Protecting OpenSSL}
\label{sec:openssl}
In this experiment, we evaluate the efficacy of our countermeasure using the attack by Shin et~al.~\cite{shin-prefetch} on the ECDH implementation in OpenSSL 1.1.0g as an example. 
Instead of re-implementing the end-to-end attack, we focus on reproducing the underlying prefetching side channel in both evaluation environments and show that \countermeasure prevents the leakage successfully.

\parhead{Vulnerability}
The leakage is caused by memory accesses to a lookup table when a point on an elliptic curve is squared. If those accesses (by chance) form a regular pattern, the prefetcher is activated and fetches memory lines before and/or after the lookup table. This leaves traces in the cache state of shared memory, leaking relations between different portions of the point on the curve. Depending on the context where this operation is used, the point may be secret information.

\parhead{Experiment}
We identify the OpenSSL library function \texttt{BN\_\-GF2m\_\-mod\_\-sqr\_\-arr} as the function that operates on the lookup table. Our test program calls this function with a value that produces a regular access pattern and thus triggers the prefetcher (if enabled). It then accesses the potentially prefetched location, in our case the first cache line after the lookup table, and measures the memory latency to determine its cache state.

We repeat the experiment in two configurations.
In the first configuration, we call the function without any countermeasure against prefetching-based side channels enabled. This experiment serves as a baseline and shows that the library function actually leaks information when called with specific inputs.
In the second configuration, we set the prefetch-disable flag before calling the library function and clear it after returning from the library function. If \countermeasure is effective, we expect no more prefetching leakage.

\begin{figure}
  \centering
  \includegraphics[scale=0.65]{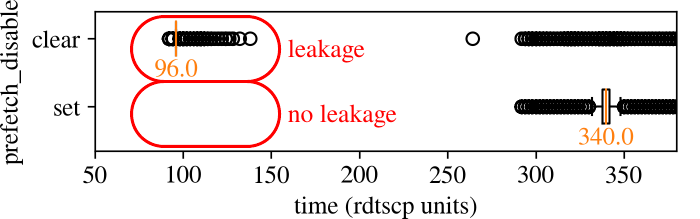}

  \textbf{(a) Intel Core i7-10510U}
  
  \includegraphics[scale=0.65]{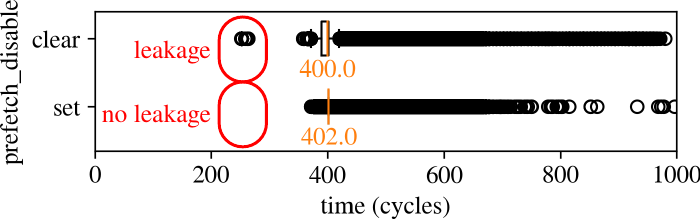}

  \textbf{(b) Raspberry Pi 4 (Broadcom BCM2711, Cortex-A72)}
  \caption{Latency of accessing the prefetch location  after calling the vulnerable OpenSSL function with the \countermeasure countermeasure not applied (\texttt{prefetch\_\-disable} flag cleared) and applied (flag set).
  Short access latency indicates unwanted leakage, which is prevented by activating our countermeasure.}
  \label{fig:openssl-effective}
\end{figure}

\parhead{Results}
We run both configurations in both evaluation environments and present the results in Figure~\ref{fig:openssl-effective}.
We repeat each configuration 1,000,000 times on the Intel CPU and 10,000,000 times on the ARM CPU. 
When the \texttt{prefetch\_\-disable} flag is cleared on the Intel CPU, we observe a significantly lower latency when loading from the memory line right after the lookup table (median: 96 units). This indicates that the prefetcher loaded this memory line into the cache (i.e. unwanted leakage).
In contrast, when \countermeasure is activated on the Intel CPU by setting the \texttt{prefetch\_\-disable} flag, the observed memory latency is above typical values for cache hits (median: 340), indicating a cache miss and absence of leakage.
On the ARM CPU, we observe a weaker leakage signal (possibly indicating that the prior work attack would not perform as well here). Without a countermeasure, the prefetching leakage is visible in the form of outliers appearing at around 250 cycles.
When we set the \texttt{prefetch\_disable} flag before calling the target function, we observe that these outliers disappear reliably.
We conclude that \countermeasure successfully prevents the prefetch-based side channel in both environments.

\subsection{Efficiency: Non-critical Workloads (Scenarios 1 and 2)}
\label{sec:non-critical-workload}

We now investigate the efficiency of \countermeasure, starting with the performance impact on workloads that are not security-critical.

\parhead{Experiment}
We run \emph{SPEC CPU 2017} benchmarks~\cite{speccpu2017} on three different system configurations.
As a baseline, we measure the performance of the SPEC workloads on a stock kernel while prefetching is either enabled or disabled permanently (scenario 1).
These measurements show us how much different workloads benefit from prefetching at all, and how expensive the radical-but-simple defense of disabling the prefetcher permanently would be.
Afterward, we measure the performance of the same workloads on a patched kernel with prefetching enabled, and without setting the \texttt{prefetch\_disable} flag (scenario 2).
This allows us to rate the performance impact on non-security-critical workloads caused by the added code that is executed on every context switch.

\parhead{Benchmark Parameters}
We run the \emph{SPEC CPU 2017 Integer Rate} set of benchmarks and report the execution time of the individual benchmarks as a metric for their performance. We run each benchmark three times (which is the maximum number of iterations in a ``reportable run''~\cite{speccpu2017:docs-iterations}).

\begin{figure}
  \includegraphics[width=\linewidth]{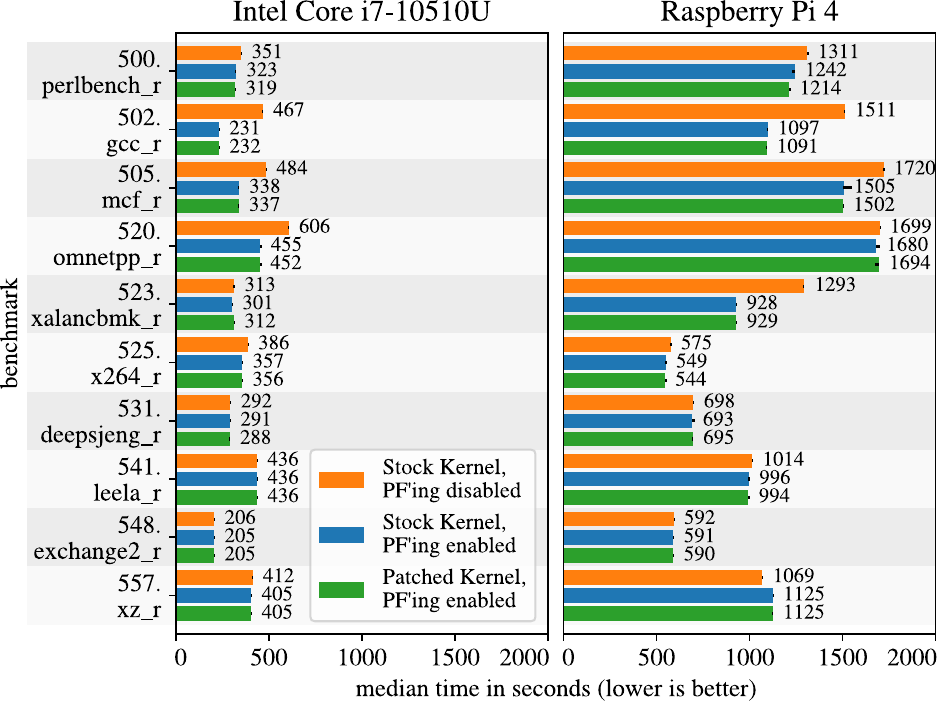}
  \caption{SPEC CPU 2017 benchmark results. Disabling the prefetcher permanently causes significant performance overhead in benchmarks 502 to 523. The performance overhead introduced by our patched kernel is negligible for non-security-critical workloads.}
  \label{fig:spec}
\end{figure}

\parhead{Results}
Figure~\ref{fig:spec} shows the benchmark results in both evaluation environments.
The bars represent the median runtime of the individual benchmarks across the three iterations, while the black error bars indicate the runtime of the other two iterations.

Comparing the two stock kernel configurations (orange and blue bars), we find that the prefetcher especially speeds up the benchmarks 502--523. At a maximum, the prefetcher improves performance by 43~\% (benchmark 505 on the Intel CPU) and 37~\% (benchmark 502 on the Raspberry Pi), respectively. In most other workloads, both configurations performed similarly. In one exceptional case, we see a slowdown by 5~\% caused by the prefetcher (557 on the Raspberry Pi).
We conclude that disabling the prefetcher permanently can lead to a significant performance drop on both tested systems.

When we compare the stock kernel and the patched kernel, both with prefetching enabled (blue and green bars), we observe only small differences in execution time. For most benchmarks, the absolute difference is around 1~\%. We conclude that our kernel patch has negligible impact on non-critical workloads.

\subsection{Efficiency: Security-critical Workloads (Scenario 3)}
\label{sec:lighttpd}

Next, we evaluate the performance impact of \countermeasure on a security-critical workload that uses our protection mechanism.

\parhead{Experiment}
To evaluate the efficiency of \countermeasure in a realistic end-to-end scenario, we now apply it to real-world software. We use the web server \emph{lighttpd 1.4.75}~\cite{lighttpd:2024} (released in March 2024) as an example.
Lighttpd ships with plugins for various cryptographic libraries that can be used as backends to provide HTTPS support. In the following experiments, we use the OpenSSL plugin for this purpose.
We compare two approaches of applying \countermeasure: on plugin level and on application level.

\parhead{\countermeasure on Plugin Level}
In this approach, we modify the Open\-SSL plugin of lighttpd such that the \texttt{pre\-fetch\_\-dis\-able} flag is set whenever the control flow enters any function in the plugin code (which then calls OpenSSL), and cleared before the control flow returns from the plugin code. This approach allows the majority of the web server code base to benefit from prefetching, but causes frequent system calls to enable or disable the prefetcher.

\parhead{\countermeasure on Application Level}
In this approach, we set the \texttt{pre\-fetch\_\-dis\-able} flag when lighttpd is started and clear it when it quits. In other words, the prefetcher is disabled for the whole lighttpd code base, including non-critical server code and any hosted web application that is executed in a child process. While this means that the server cannot benefit from prefetching at all, it also means that fewer system calls are required.

\parhead{Web Application}
In our example, we use lighttpd to host a web application that we expect to benefit from prefetching in a scalable way. Our example application is written in Python and attached to lighttpd via CGI. It accesses a table in an SQLite database filled with random numbers. We set up the table with 16 columns and a variable number of rows. Whenever the application is called, it randomly selects one of the columns and finds the minimum value in that column. Thus, the application needs to access every value in that column. We expect that this activity can benefit from prefetching, especially for tables with many rows.

\parhead{Benchmarking Approach}
We configure lighttpd to serve our web application via HTTPS over TLSv1.3.
To measure the server's performance, we use the \emph{httpit} HTTP(S) benchmark~\cite{httpit} over the local loopback interface.
We configure httpit to perform one connection at a time and measure the average number of requests per second that the server is able to serve over a 5-minute time period as a performance metric.

To rate the performance of this setup when protected using \countermeasure, we measure it in scenario 3, i.e., with a patched kernel and prefetching temporarily disabled.
For completeness and as a baseline to compare with, we also include performance measurements of the same setup in scenario 1 (stock kernel) and scenario 2 (patched kernel with prefetching permanently enabled). 

\begin{figure}
  \includegraphics[width=\linewidth]{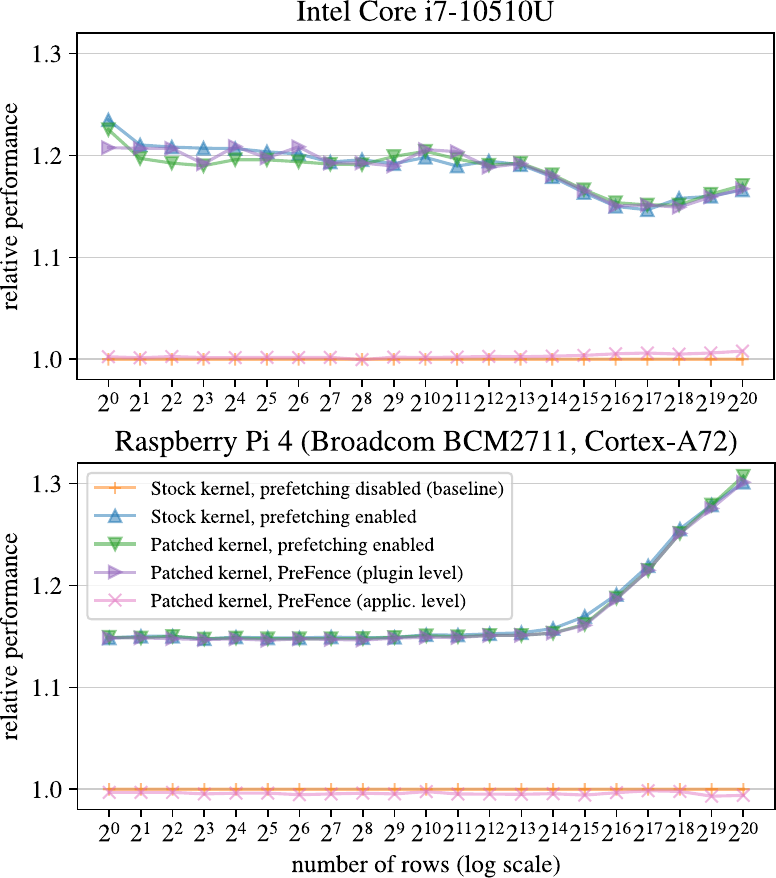}
  \caption{
    Lighttpd benchmark results.
    Applying \countermeasure on plugin level results in performance similar to permanently enabled prefetching.
    Applying \countermeasure on application level results in performance similar to permanently disabled prefetching.
  }
  \label{fig:lighttpd-progressive}
\end{figure}

\parhead{Results: Plugin-level vs. Application-level Defense}
We present the results of this experiment in Figure~\ref{fig:lighttpd-progressive}.
The y-axis represents the performance of the web server relative to a baseline. This baseline is the performance of the web server in scenario 1 (stock kernel) with prefetching permanently disabled, indicated by the orange line in the plots.

We find that applying \countermeasure on plugin level (purple line) results in negligible performance impact of less than 1~\% on average compared to those scenarios where prefetching is permanently enabled, either on the stock kernel or on the patched kernel (blue and green lines, respectively). Thus, the performance impact of \countermeasure is negligible in this case.
In contrast, we find that applying \countermeasure to the whole web server (pink line) results in a performance roughly equal to the baseline, i.e., turning off prefetching permanently.
Therefore, we conclude that \countermeasure is best used on a fine-grained level if possible, as this allows most parts of the application to still benefit from prefetching and keep a high performance level.

\parhead{Results: Platform-specific Impact of SQL Query Complexity}
On the x-axis of Figure~\ref{fig:lighttpd-progressive}, we show the number of table rows traversed by the web application.
We find that enabling prefetching for the runtime of the web application always improves its performance, even if the table contains only a single row.
Because the prefetcher is unlikely to speed up the actual database lookup in a table with only one row, we attribute the speedup in that case to beneficial prefetching activity in code that is not directly related to the database lookup (such as the Python interpreter or the CGI handling routines in the web server).
We further note that the relative performance compared to the baseline increases for higher row counts on the Raspberry Pi. This indicates that also the actual database lookup benefits from prefetching on that platform, at least for higher row counts.
However, this is not the case on the Intel CPU, which indicates that the prefetcher on that CPU generally does not speed up this workload. As pointed out by prior work, prefetching on the Raspberry Pi is more aggressive than on the Intel CPU~\cite{fetchbench}, which likely explains this discrepancy between the two architectures.

\subsection{Summary of Results}
We have demonstrated that \countermeasure is applicable to our Intel and ARM processors, as their prefetchers cannot be trained while they are disabled (Section~\ref{sec:disabled-behavior}).
We also verified that \countermeasure is effective, as it eliminates leakage caused by the prefetcher in a known-vulnerable library function on both architectures (Section~\ref{sec:openssl}).
We further demonstrated that \countermeasure is efficient:
For workloads that are not security-critical, such as SPEC benchmarks, we observed a performance overhead of less than 1~\% for the majority of the benchmarks (Section~\ref{sec:non-critical-workload}).
For our example of a security-critical workload, a lighttpd web server hosting a web application that may benefit from prefetching, we found that the performance overhead is negligible (i.e., less than 1~\% on average) if \countermeasure is applied on the level of the OpenSSL plugin. In this way, the prefetcher is only disabled when security-critical parts of the code are executed (Section~\ref{sec:lighttpd}).

\section{Prior-Work Software-Based Countermeasures}
\label{sec:prior-countermeasures}

We now answer RQ\ref{rq:mitigations} by discussing prior work mitigations to pre\-fetch-based attacks. We show that none of them provide an easy and efficient way to prevent attacks.
We classify a countermeasure as \emph{software-based} if it can be applied to at least one real-world CPU architecture via (modified) software.

\subsection{Prior Work Countermeasures}

\parhead{Constant-Time Programming~\cite{fetchbench,shin-prefetch,afterimage,xiao:2023}}
One way to prevent most pre\-fetching\--based side channels---and also other cache-timing side channels---is the programming technique of \emph{constant-time programming}.
Despite the name, it refers not only to writing code that executes in the same time regardless of the (potentially secret) information processed, but also mandates that no secret-dependent control flow or memory access patterns occur~\cite{intel-constant-time}.
Avoiding secret-dependent memory access patterns prevents address-based prefetchers from transferring secrets into their internal state during training. Moreover, avoiding secret-dependent branches prevents prefetcher-based attacks that infer the victim's control flow based on conditionally executed load instructions.
However, a recent work has shown that constant-time programming is ineffective against DMP-based attacks that exploit secret-dependent \emph{values} instead of addresses or branches~\cite{chen:2024:gofetch}. In addition, making code constant-time requires complex re-writes and results in significantly reduced performance~\cite{pereidagarcia:2016:consttime,cauligi:2020:consttime}.

\parhead{Clearing the Prefetcher's State on Context Switches~\cite{fetchbench,afterimage,cronin:2019}}
Some prefetching-based attacks train the prefetcher in a victim context, then switch into the attacker's context and trigger it there. In practice, this context switch can be a switch between two userspace processes, from kernel to userspace, or a return from trusted execution. To mitigate such attacks, the prefetcher's state could be cleared on context switches.
On Linux systems, switches between kernel and userspace and within userspace are handled by the scheduler. As processes cannot control when the scheduler interrupts (or resumes) them, clearing the prefetcher's state needs to be implemented in the scheduler.
For context switches between trusted execution and operating system, the prefetcher could be cleared before returning control from the TEE back to the (untrusted) operating system.
The process of clearing the prefetcher's state is straightforward to implement if the CPU provides a suitable instruction, such as \texttt{CPP RCTX} on some ARM CPUs~\cite{arm:2020:prefetch-flush-instr,arm:2023:prefetch-asa}. Otherwise, all patterns need to be evicted from the state and replaced with secret-independent ones. This process is non-trivial to implement, as it requires knowledge of implementation details such as the number of patterns stored and the replacement policy.

We note that clearing the prefetcher's state on context switches is an incomplete countermeasure in three cases.
First, it is not applicable to attacks that trigger the prefetcher in the victim process.
Second, this countermeasure assumes that the prefetcher is not shared across physical or SMT sibling cores. Otherwise, the attack could be executed from a different core before the context switch resets the state.
Third, in the case of trusted execution, prior work has shown that an attacker is able to interrupt trusted execution before it completes~\cite{sgxstep,ryan:2019:interrupt-tz,kou:2021:loadstep}. If this is possible, clearing the prefetcher's state only at the end of a trusted execution procedure is insufficient.

\parhead{Mitigating Cache-Timing Side Channels~\cite{shin-prefetch,afterimage}}
We found in Section~\ref{sec:attacks-conclusion} that all prefetching attacks rely on a cache-timing side channel. To mitigate those, access to timer interfaces can be restricted or their resolution can be reduced. This mitigation has especially been applied to browsers in the past~\cite{chromium-timer-precision,mozilla-timer-precision}.
However, modern browsers provide many indirect ways to acquire timestamps~\cite{rokicki:2021:jstimers} and also in other contexts, attackers may fall back to alternatives such as a counter thread as a timer replacement~\cite{armageddon}.
General countermeasures against cache-based side channels have been discussed in prior work extensively~\cite{osvik:2006,flush-reload,zhang:2013:dueppel,liu:2016:catalyst}, but none were implemented on a large scale.
Consequently, we consider it next to impossible to reliably block an attacker from all possible ways to generate precise timestamps.

\parhead{Anomaly Detection~\cite{afterimage}}
During a prefetcher-based side channel attack, the attacker may execute code that results in unusual amounts of cache flushes or evictions, leading to unusual amounts of cache hits and misses. In addition, if the attacker interacts a lot with the prefetcher in order to prime or trigger it, the amount of prefetch-related events may increase. If the CPU exposes performance counters for these events, those can be observed and evaluated to detect unusual activities~\cite{payer:2016:hexpads,zhang:2016:cloudradar}.
However, such a heuristic detection system will produce false-positive alerts, miss malicious events (false negatives), and introduce a constant runtime overhead affecting all workloads. In addition, intrusion detection systems generally do not prevent attacks, but merely detect them and react after the attack has started. Thus, we consider this mitigation strategy incomplete.

\parhead{Security-Aware Core Assignment~\cite{augury,chen:2024:gofetch}}
If a prefetcher operates per-core, one possible mitigation is more advanced core assignment.
For processors with a heterogeneous design, a vulnerable prefetcher may only be present on some of the cores. In this case, security-critical or untrusted workloads could be assigned to cores that are not vulnerable.
Similarly, a vulnerable per-core prefetcher could be disabled on one of the cores. This core could then be reserved for critical workloads.
However, it is not trivial in practice to decide which processes can be assigned to which core.
In addition, reserving a core for critical operations is likely to reduce overall system performance significantly: If critical workloads are frequent or long-running, assigning them to a single core will limit their throughput. If critical workloads are rare or short-running, reserving one core for them will result in the core idling most of the time.

\parhead{Oblivious Execution~\cite{afterimage}}
The idea behind oblivious execution~\cite{rane:2015} is to eliminate the side-channel effect of a secret-dependent if/else conditional by executing both branches and only persisting the result of the correct branch in memory in the end. In this way, the attacker is no longer able to distinguish both branches: the timing and the memory patterns that are executed are always the same (as long as the branches do not contain more secret-dependent instructions).
However, this approach obviously comes with a significant performance overhead. Rane et~al.~\cite{rane:2015} report a significant mean overhead of $16.1\times$.

\parhead{Blinding~\cite{chen:2024:gofetch}}
A DMP may be triggered by a data value that matches a specific pattern, e.g., that looks like a pointer. To ensure that such a prefetcher is not triggered by an untrusted value, a mask can be added to the value before it is stored in memory and removed after it is loaded again.
However, implementing this countermeasure is not trivial and introduces memory and computational overhead.

\parhead{Disabling the Prefetcher~\cite{fetchbench,shin-prefetch,afterimage,cronin:2019,chen:2024:prefetchx,chen:2024:gofetch}}
If the CPU exposes a way to control the prefetcher, the most straightforward way to prevent any leakage from the prefetcher is to disable it permanently. Some CPUs expose such configuration options through model-specific registers (MSRs).
However, this countermeasure comes with a significant performance decline for workloads that benefit from prefetching, as we show in Section~\ref{sec:non-critical-workload}.

While the general idea of disabling the prefetcher temporarily has been stated in literature before~\cite{fetchbench,chen:2024:gofetch}, no detailed design, implementation, or evaluation has been provided to date. We fill this gap by presenting \countermeasure in this paper. Our novel design has low runtime overhead and handles the special case of prefetcher sharing across cores or SMT siblings by extending the process scheduler.

\subsection{Conclusion: Prior-Work Countermeasures Are Costly or Incomplete}
In summary, every prior-work countermeasure violates one of our design goals stated in Section~\ref{sec:design-goals}. We consider
\emph{constant-time programming} complex to implement, expensive at runtime, and ineffective against DMP-based side channels;
\emph{clearing the prefetcher's state on context switches} specific to attacks that trigger the prefetcher in the attacker's context, expensive at runtime, and incomplete when using SMT;
\emph{mitigation of timing sources} and \emph{anomaly detection} inherently incomplete approaches;
\emph{security-aware scheduling} hard to implement in an efficient way;
\emph{blinding} complex to implement, expensive at runtime, and specific to DMP-based side channels;
\emph{oblivious execution} and \emph{disabling the prefetcher permanently} expensive at runtime.

\section{Discussion}

\subsection{Applicability to Different Scopes}
\label{sec:discussion:scope}
Our current implementation of \countermeasure, as presented in Section~\ref{sec:impl}, protects a userspace process from attacks by other userspace processes (scopes SP, CT, and CP). This covers 11 out of 13 attacks that we discussed in Section~\ref{sec:systematization}.
The general principle of \countermeasure can also be used to protect kernel code from attacks from userspace (scope KU): the kernel could disable the prefetcher temporarily during security-critical operations.
To prevent attacks from an untrusted operating system on a trusted execution environment (scope TO), the prefetcher could be disabled temporarily while security-critical code is executed in trusted execution.
However, because the untrusted operating system can usually interrupt trusted execution~\cite{sgxstep,kou:2021:loadstep,ryan:2019:interrupt-tz}, the prefetcher's activation state needs to be saved when an interrupt causes a context switch to the untrusted operating system and restored when switching back. In the case of Intel SGX, which is interrupt-unaware~\cite{sgxstep,intel-sgx-paper}, this would require additional hardware support.

\subsection{Applicability to Other Architectures And Flags}
\parhead{Other Architectures}
In this paper, we implement and evaluate \countermeasure for two specific CPUs. Our targets are chosen to represent popular attack targets (such as the Intel IP stride prefetcher~\cite{shin-prefetch,afterimage,xiao:2023}) and cover two popular architectures (x86\_64 and ARMv8). However, prior work has revealed that prefetcher implementations can differ widely, even across processors of the same brand or architecture~\cite{fetchbench}.
We are confident that the design of \countermeasure is general enough to be transferred to other prefetchers as well, as long as those fulfill two basic requirements.
First, the processor must expose a way to disable the prefetcher dynamically at runtime, for example by setting a bit in an MSR.
Second, the prefetcher must not update its state while disabled.
Unfortunately, there are some prefetcher implementations that do not fulfill the first requirement.
For instance, Intel does not publicly disclose a way to control the XPT prefetcher, even though this prefetcher can be controlled from BIOS configuration tools~\cite{intel:2023:xpt-control}.
Similarly, there is no publicly known way to disable the DMP prefetcher on Apple M1 and M2 processors~\cite{chen:2024:gofetch}.
Thus, we strongly encourage processor manufacturers to add such flags to future processor designs and ideally establish a well-documented standard to control prefetchers.

\parhead{Other Flags}
We emphasize that \countermeasure's general design is not limited to managing prefetching-related flags. \countermeasure could be adapted to control other microarchitectural components during the execution of security-critical code as well.
For instance, ARM and Intel recently introduced the \emph{Data-Independent Timing (DIT) flag}~\cite{arm:2024:arch-regs,apple:2024:dit} and the \emph{Data Operand Independent Timing Mode (DOITM) flag}~\cite{intel:2023:doit}, respectively. Newer processors only guarantee data-independent timing behavior for certain instructions when these flags are set. These flags are meant to be set while cryptographic operations implemented in constant-time code are executed. They temporarily disable a range of optimizations that impact timing behavior (including, but not limited to certain prefetchers~\cite{chen:2024:gofetch,intel:2023:doit}). \countermeasure could be adapted to dynamically set and clear these (or other) flags as well.

\subsection{Covert Channels}
In this paper, we focus on mitigating prefetching side channels and exclude covert channels from the scope.

Six prefetching-based covert channels have been proposed in recent work.
Cronin et~al.~\cite{cronin:2019} implement a covert channel using the Intel IP stride prefetcher. They prime the prefetcher from the receiver's end, either evict or keep the primed patterns from the sender process, and finally probe for the existence of the primed patterns in the receiver process. This probing either triggers the prefetcher or not, indicating either a 0-bit or a 1-bit, respectively.
Chen et~al.~\cite{afterimage} exploit the same prefetcher but encode the information to transmit into the stride.
Rohan et~al.~\cite{rohan:2020} exploit the Intel stream prefetcher. The sender triggers the prefetcher on shared memory. The direction of prefetch (forward or backward) is interpreted as a 1-bit value by the receiver.
Schlüter et~al.~\cite{fetchbench} encode bit vectors into the region-based ARM SMS prefetcher to transfer information from trusted execution to the untrusted OS.
Chen et~al.~\cite{chen:2024:prefetchx} implement two covert channels based on the Intel XPT prefetcher. In the first scenario, the receiver primes the prefetcher's state and the sender either idles or changes the state by evicting one of the primed entries. In the second scenario, the sender either trains or resets the prefetcher for a shared page.

All of these attacks have in common that not the victim but the attacker controls the training stage. In fact, there is no victim process at all in a typical covert-channel setting: Covert channels are incapable of leaking secret information on their own. Rather, they are a means of exfiltrating secret information that the attacker has obtained in another way beforehand.
As these attacks do not leak information out of a victim process directly, a victim process has no incentive to defend against them. Thus, \countermeasure is not applicable.

\subsection{Hardware-Based Countermeasures}
In this paper, we focus on software-based mitigations because they can easily be applied and evaluated on current hardware. However, we also want to briefly discuss countermeasures against prefetching side channels that require hardware modifications.

\parhead{Choosing a different trigger~\cite{fetchbench}}
Some attacks rely on collisions on the prefetch trigger, e.g., the address of a load instruction. To mitigate such attacks, such collisions should be avoided. For example, the IP stride prefetcher of many Intel CPUs identifies a stride pattern by the partial address of a load instruction that caused the memory access. Because only few bits of that instruction address are considered, an attacker can cause such collisions on purpose.

The first intuition to fix this issue is to use full instruction addresses instead of partial addresses. However, this does not rule out the possibility of collisions completely, for instance in case of a shared library.
To avoid this problem, the instruction address must not be used as the only trigger. For example, a process identifier could be added. Only if the process identifier \emph{and} the instruction address match, a prefetch can be triggered.

\parhead{Partitioning the Prefetcher's State~\cite{cronin:2019,afterimage,fetchbench,chen:2024:prefetchx}}
Similar to adding a process identifier, the prefetcher's state table could be partitioned based on one of multiple criteria. To avoid leakage between different privilege levels, such as kernel and userspace, or trusted execution and untrusted OS, the prefetcher could keep track of the privilege level that a pattern belongs to.

However, merely keeping track of the privilege level does not protect against attacks within the same level, such as leakage between two userspace processes. To separate those, it is again necessary to store an additional process identifier.

Partitioning protects from triggering a prefetch pattern accidentally in a wrong context. This ensures that no information about the stored patterns is leaked across privilege domains. However, depending on the implementation, attackers may still be able to prime the prefetcher with attacker-controlled patterns that are then potentially evicted by victim activity, leaking control flow information.

\parhead{Extending the Instruction Set~\cite{augury,fetchbench}}
Another option is to extend the instruction set by instructions that change the pre\-fet\-cher's state or behavior.
First, an instruction could be added that flushes the prefetcher's state. This instruction could then be called by the operating system on context switches, or when switching between privilege domains. However, this approach only works when the prefetcher is not shared among multiple cores or SMT threads.
Second, a special load instruction to be used in security-critical code sections could be added that does not influence the prefetcher's state.

\section{Conclusion}
\label{sec:conclusion}

In this work, we addressed the challenge of efficient defenses against side-channel attacks exploiting the prefetcher to leak secret information from a victim userspace process. We started by providing the first systematic analysis of the existing \numAttacks related attacks from literature, and showed that all of them rely on three main stages: prefetcher training, prefetcher triggering, and cache extraction.

Our proposed countermeasure, \countermeasure, allows vulnerable programs to ensure that they do not train the prefetcher. More precisely, it enables processes to selectively disable the prefetcher from userspace, while ensuring that parallel processes sharing the same prefetcher on other cores or SMT siblings are considered as well. In addition, issues in re-scheduling of processes are considered, and handled transparently for the vulnerable process. \countermeasure enables fine-grained control to minimize the time the prefetcher is unavailable, while ensuring that the critical prefetcher training attack stage cannot target the victim process any longer.
Our prototype implementation of \countermeasure for Intel x86\_64 and ARM Cortex-A72 processors is a Linux kernel patch to the scheduler to automatically ensure correct prefetcher state on re-scheduling of processes, together with a system call for processes to directly enable or disable the prefetcher. 
We provide our \countermeasure implementation as open-source software.

We demonstrated the efficacy of our approach by preventing a prior work attack successfully.
Interestingly, our performance evaluation showed that the performance overhead is negligible if only parts of the application execute security-relevant code. 
In particular, we showed that the performance impact of our solution on non-security-critical code is usually below 1~\%.
For security-critical workloads, we found its performance impact to be dependent on the way \countermeasure is applied to the code; for a real-world web server application, we showed that the overall performance impact is negligible (less than 1~\% on average) when \countermeasure is applied only to code sections related to cryptographic operations.

In conclusion, we presented a user-friendly and efficient sched\-uling-aware countermeasure to protect victim processes against prefetcher side channels, founded on a systematic analysis of prior work attacks and countermeasures. 
We expect our countermeasure could widely be integrated in commodity OS, and even be extended to signal generally security-relevant code to the kernel to allow coordinated application of countermeasures (e.g., DIT flags).

\bibliographystyle{ACM-Reference-Format}
\bibliography{paper}

\appendix

\section{Prior-Work Prefetching-based Attacks}
\label{sec:attacks-detail}
In this section, we give a high-level overview of prefetching-based attacks in prior work and explain our dissection of these attacks into the five stages as illustrated in Figure~\ref{fig:attacks-flow}.

\parhead{Shin et~al.~\cite{shin-prefetch}}
This paper exploits the Intel IP stride prefetcher and targets the ECDH implementation in OpenSSL. In the preparation phase (S1), the attacker identifies memory lines the OpenSSL shared library code that are cached only conditionally depending on the supplied input. These cache lines only appear in cache when an internal lookup table is accessed such that a sequence of accesses forms a regular stride pattern. In the reset phase (S2), these memory locations are flushed from the cache by the attacker. Then, the attacker calls the library. The library trains (S3) and triggers (S4) the prefetcher only if the supplied input leads to memory loads that form a stride pattern. The attacker probes the cache lines through a cache-timing side channel (S5) to decide whether the input triggered the prefetcher or not. This prefetching-based primitive is then embedded in a differential attack to recover the secret.

\parhead{Augury~\cite{augury}}
This paper is the first to investigates the data memory-dependent prefetcher (DMP) of the Apple M1 SoC. It describes the prefetcher's behavior when multiple pointers that are stored sequentially in memory (e.g., in an array) are loaded and dereferenced: The DMP fetches subsequent pointers and dereferences them.
The authors present multiple approaches to exploit this behavior.

The first approach (``Out-of-bounds reads'', \emph{Augury OOB}) assumes that a user secretly selects a pointer from a finite list of candidate pointers. This pointer is stored just behind an array of pointers in the victim's memory. The goal of the attacker is to find the chosen pointer without accessing it architecturally. The attack is described with Flush+Reload and Prime+Probe; we discuss the more complex Prime+Probe variant.
The attacker sets up an eviction set for each of the candidate pointers (S1) and loads them (S2). Next, all the pointers in the pointer array are dereferenced to train the prefetcher (S3). When the end of the array is reached, the prefetcher is triggered to prefetch past the array bound (S4) and to dereference the user-chosen pointer. By timing the access latency to all candidate pointers using the eviction sets (S5), the attacker decides which of the pointers was chosen.

In a second approach (\emph{Augury SLH}), the authors discuss the impact of the DMP on code that uses speculative load hardening (SLH)~\cite{llvm-slh} to protect against Spectre attacks~\cite{spectre}. 
The core idea of SLH is to verify untrusted (e.g., user-supplied) memory offsets during speculative execution to prevent speculative out-of-bounds accesses. The compiler adds branchless code that replaces out-of-bounds offsets with a safe value (often 0) using binary arithmetic.
In Augury's example, a code snippet trains the prefetcher by iterating over a pointer array (S3). While SLH prevents \emph{speculative} out-of-bounds reads, the prefetcher is still able to prefetch past the array bound when triggered by an access to the last array element (S4). Thus, a pointer that is stored just behind the pointer array can be fetched and dereferenced by the prefetcher, leaving traces in the cache that can be recovered (S5).

A third approach (\emph{Augury Addr}) describes how the DMP can be used to determine whether an address is a valid (mapped) virtual memory address or not. To this end, the attacker sets up an array of 3 pointers, where the third pointer is the address to test (S1). The attacker ensures that the array is not cached (S2). Next, the attacker traverses the array in speculative execution and within the context where the mapping is to be checked. This trains the prefetcher (S3). As the DMP requires at least three valid addresses to be triggered for the first time (S4), prefetching will only happen if the address is valid. The attacker tests the cache state of the first out-of-bounds element after the array of pointers (S5). If this element is cached, the tested address is valid, otherwise, it is invalid.

\parhead{AfterImage~\cite{afterimage}}
This work exploits the Intel IP stride prefetcher five different ways.
Generally, these approaches exploit collisions on the instruction pointer (IP) address. The exploited prefetcher identifies patterns stored in its internal state based on the instruction address of the load instruction that caused the load. However, this instruction address is internally truncated to the 8 least-significant bits. Consequently, the attacker can cause a collision by aligning a load instruction in their own code such that its 8 least significant bits match the respective bits of the instruction address in the victim code. The prefetcher is then unable to distinguish those two instructions from different contexts.

In \emph{AfterImage variant 1}, the prefetcher is used to leak the control flow of a victim process.
This attack is described with same-process and cross-process scope as well as using Flush+Reload and Prime+Probe for extraction. We focus on the cross-process, Prime+Probe variant, which we consider the more complex one.
The attacker's goal is to determine whether a branch in a victim process is taken or not taken. To this end, the attacker selects one load instruction from each of the two potential code flows in the victim process and aligns two load instructions in their own process to them (S1). The attacker further prepares (S1) and loads (S2) eviction sets on the load targets in the victim process. The attacker then primes the prefetcher (S3) by training it in the attacker's process.
Then, a context switch to the victim process happens, where the branch is either taken or not taken and the respective load instruction is executed, further (mis)training the prefetcher (S3). As the prefetcher was pre-trained, this load will further trigger prefetching after the load target of the victim instruction (S4). The attacker extracts those prefetching effects from the cache by reloading the eviction sets (S5).

\emph{AfterImage variant 2} exploits the stride prefetcher to determine whether a branch is taken in the kernel. The authors attack a system call handler that operates on a memory buffer passed in from userspace. The idea is similar to variant 1, however, aligning to a load instruction in the kernel is more difficult as the instruction address is unknown. For this reason, the attacker first determines the offset by testing all $2^8=256$ possibilities in a process called \emph{IP matching} (S1). Then, the prefetcher is primed in the attacker process (S3). The system call is issued. If the branch is taken, the prefetcher will be further (mis)trained (S3) and triggered (S4) to prefetch memory from the buffer. After returning from the system call, the attacker probes the cache state of the respective locations in shared memory (S5).

In addition, AfterImage describes an attack on \emph{SGX}. This attack does not exploit a collision. The goal is to leak control flow from an enclave. In the described setting, a load instruction is executed in a loop within the victim enclave. The instruction loads with a secret-dependent stride from a shared buffer that is passed from userspace into the enclave. This loop trains (S3) and triggers (S4) the prefetcher. After returning to userspace, the attacker process recovers the stride by inspecting the cache state of the shared buffer (S5).

The paper further presents an attack on the \emph{RSA} implementation of MbedTLS. The victim code contains secret-dependent branches that the attacker wants to track. To this end, the attacker first identifies suitable load instructions to align to by reverse-engineering the victim binary (S1). Next, the attacker primes the prefetcher in their own memory to a high confidence level (S3) and switches to the victim code. The victim executes a colliding load instruction and re-trains the prefetcher (S3). As the loaded address will likely not match the previously trained stride, the confidence will be lowered. After switching back to the attacker process, the attacker tries to trigger the prefetcher in their own memory again (S4) and extract the prefetcher's behavior from the cache state (S5). The attacker will only observe prefetching effects if the confidence was not lowered by the victim, i.e., if the monitored branch was not taken.

Finally, the authors present a prefetcher-based \emph{synchronization primitive} that operates similar to the RSA attack. They envision this primitive could be used as a trigger for a power-based side-channel attack, for example to detect the beginning of a cryptographic operation.
Again, the attacker begins by identifying a load instruction to align to in the victim code (S1). The prefetcher is then primed on a colliding load instruction in the attacker's process to a high confidence level (S3). The attacker now switches frequently between victim and attacker code. As soon as the victim executes the target instruction, the prefetcher will be trained (S3) and the confidence will be lowered. The attacker tries to trigger the prefetcher (S4) and inspects the prefetcher's activity in the cache (S5). Once the prefetcher can no longer be triggered, the attacker knows that the victim executed the target instruction. In that case, the attacker raises a trigger signal.

\parhead{Xiao et~al.~\cite{xiao:2023}}
This paper uses the Intel IP stride prefetcher to attack an (undisclosed) AES implementation. To this end, the attacker monitors the cache state of the two memory lines just before and after the S-box. Depending on the access pattern to the S-box, which depends on plaintext and key, different prefetching activity in those cache lines is to be expected.
After identifying the cache lines to monitor (S1), the attacker flushes them (S2). Then, the encryption is performed, potentially training (S3) and triggering (S4) the prefetcher. Finally, the cache state is inspected using a timing-based side channel (S5).

\parhead{FetchBench~\cite{fetchbench}}
This paper exploits the Spatial Memory Streaming (SMS) prefetcher in the ARM Cortex-A72 processor to attack the T-table-based AES implementation of MbedTLS. The attacker's goal is to extract the encryption key from the victim process. To this end, an instruction address collision is exploited. The authors first align load instructions in the attacker process with those in the victim process that load secret-dependent values (S1). They further identify a cache line that is accessed shortly before the victim process processes the secret (S1). This line, as well as a local probe array in the attacker process, are then  flushed (S2). Next, the victim process encrypts an attacker-supplied plaintext using its own secret key. During encryption, the victim accesses multiple elements of a lookup table. The accesses to the lookup table train the prefetcher (S3). The attacker then switches into their own process using an inter-processor interrupt and triggers the prefetcher there (S4), making the prefetcher transfer the pattern learned in the victim context into the attacker's context. Finally, the attacker deduces the prefetcher's activity from the cache state (S5).

\parhead{PrefetchX~\cite{chen:2024:prefetchx}}
This paper uses the Intel eXtended Prediction Table (XPT) prefetcher to exploit the RSA implementations of MbedTLS and GnuPG. The XPT prefetcher is the only known prefetcher that is attached to the last-level cache (LLC) and thus shared across cores.
It keeps a list of recently accessed pages and counts the number of cache misses per page. Once such a miss counter surpasses a fixed threshold for a page, the prefetcher effectively bypasses the LLC for future loads from that page.

The attacker and victim processes run on different cores. To start from a clean cache state, the attacker sends a signal to the victim process (S2). This enforces a context switch into the kernel and the resulting overhead evicts the target cache line. Next, the attacker fills the prefetcher's state with pages that they control (S3). Then, the victim executes a code section containing a secret-dependent load. If the load is performed, the prefetcher's state is updated and one of the attacker's pages are evicted from the state (S3). The attacker then checks the prefetcher's state by triggering it on their own pages (S4) and measures whether the prefetcher still triggers or not (S5).

\parhead{GoFetch~\cite{chen:2024:gofetch}}
This paper revisits the DMP prefetcher of the Apple M1 SoC and its successors and discovers that the DMP is more than a pointer array prefetcher: It can be triggered by merely loading a single address from memory, even without dereferencing it.

The paper presents attacks on four real-world targets: the Go implementation of the RSA cryptosytem, the OpenSSL implementation of the Diffie-Hellman key exchange, and implementations of the post-quantum algorithms CRYSTALS-Kyber and CRYSTALS-Dilithium.
All attacks are based on the same idea. The attacker crafts malicious inputs that, when combined with secrets during computation of the target algorithm, result in intermediate values that form a valid pointer if and only if the secret fulfills a certain condition. The prefetcher is only activated if the condition is fulfilled, leaking information about the secret.
On a high level, those attacks perform the following steps.
First, an attacker process prepares (S1) and loads (S2) eviction sets that evict the pointer's anticipated location and target address.
Then, the input is crafted and supplied to the victim process. If the condition is fulfilled, an intermediate value in the victim context forms a pointer. Once this pointer is loaded, the prefetcher is trained (S3, it updates its history) and triggered (S4, it dereferences the pointer).
Finally, the attacker process re-loads the eviction sets (S5) to determine whether the condition was fulfilled or not.

Notably, these attacks apply even to constant-time implementations, which only guarantee constant execution time and (architectural) memory access locations, but do not constrain intermediate values.

\section{\countermeasure Efficiency Evaluation: Overhead on Context Switch and System Call}
\label{sec:appendix-eval}

In this section, we perform two additional experiments on the efficiency of our \countermeasure implementation. We measure the fixed overhead caused by the additional kernel code that needs to be executed on every context switch and the overhead of performing a system call in order to set or clear the \texttt{prefetch\_disable} bit of a process. Both experiments were only performed on the Intel CPU.

\subsection{Fixed Overhead on Context Switch}
\label{sec:overhead-context}
\parhead{Experiment}
In this section, we evaluate the fixed overhead that our \countermeasure implementation adds to every context switch.
We first measure the execution time of a context switch in the stock kernel and compare it to the execution time in our patched kernel.

To this end, we implement two userspace processes that share a memory page.
Both processes are pinned to the same CPU core.
The first process constantly writes the current high-resolution timer value retrieved from the \texttt{rdtscp} instruction to shared memory. The second process reads the timer value from shared memory and computes the difference to the current timestamp.
When the first process is scheduled, it keeps incrementing the timestamp written to memory until it is descheduled. Next, the second process is scheduled, computes the timestamp difference and logs the result.
We filter out ``zero samples'' caused by the second process being re-scheduled before the first one, i.e., where the timestamp in memory has not been incremented compared to the last execution of the second process.
We run this experiment on an idle system to maximize the probability of a context switch between our two processes.

\begin{figure}
  \centering
  \includegraphics[scale=0.65]{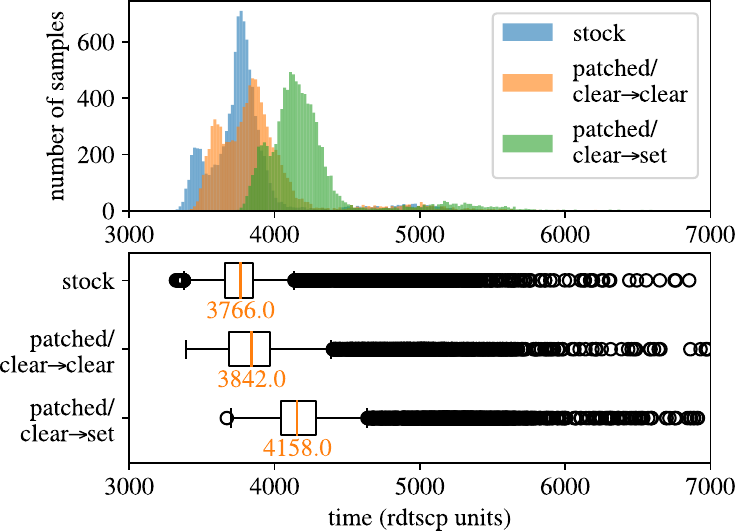}
  \caption{Context switch overhead on the stock kernel and on our patched kernel. For the patched kernel, we evaluate transitions between normal processes (\texttt{prefetch\_\-disable} bits cleared) and the transition into a security-critical process (from bit cleared to bit set). The added overhead is negligible.}
  \label{fig:context-switch}
\end{figure}

\parhead{Results}
We measure the execution time of a context switch in three scenarios on our Intel CPU:
(1) with the stock kernel,
(2) with our patched kernel, switching between two processes with the \texttt{prefetch\_\-disable} bit cleared,
(3) with our patched kernel, switching from a process with the \texttt{prefetch\_\-disable} bit cleared into a process with the bit set.
We repeat each experiment until 10,000 non-zero samples have been collected.

We present our results in Figure~\ref{fig:context-switch}. Not surprisingly, the stock kernel has the smallest median context switch execution time of 3766 time units. Switching between two normal processes on our patched kernel requires 3842 time units (median), an negligible increase of 76 units or 2~\%. When the prefetcher's state needs to be changed on context switch, the median execution time is 4158 units, an increase of 392 units or 10~\% compared to the stock kernel.

\subsection{One-off Overhead of the System Call}
\label{sec:overhead-syscall}
\parhead{Experiment}
We set up a userspace process that performs the \texttt{prctl} system call twice, once to set the \texttt{prefetch\_\-disable} bit, and once to clear it again. Before and after each of the system calls, we use the \texttt{rdtscp} instruction to get high-precision timestamps. Finally, we compute the difference between the timestamps.

\begin{figure}
  \centering
  \includegraphics[scale=0.65]{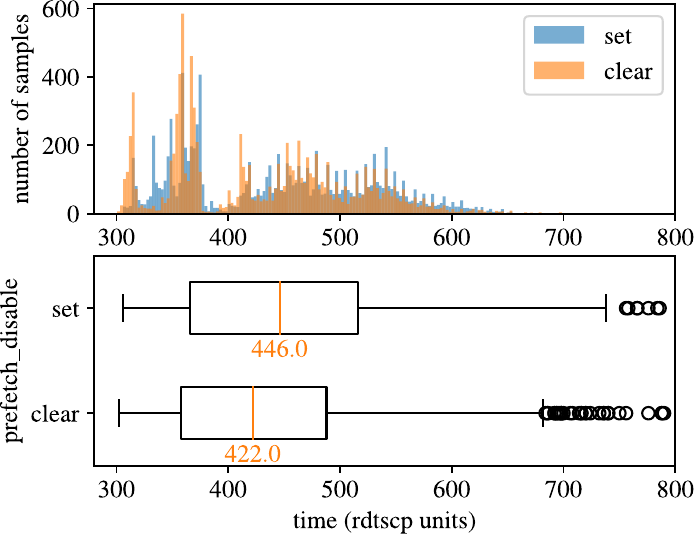}
  \caption{Overhead of a system call that sets or clears the \texttt{prefetch\_\-disable} bit.}
  \label{fig:syscall}
\end{figure}

\parhead{Results}
We repeat the experiment 10,000 times on our Intel CPU.
The results are plotted in Figure~\ref{fig:syscall}.
We note that the median duration for both system calls is around 430 units.
The median duration of the system call to set the flag is negligibly longer than the median duration of the system call to clear it.
For comparison, the overhead of the system call is roughly in the same order of magnitude as a memory load that misses the cache (in the OpenSSL experiment in Section~\ref{sec:openssl}, we observed that a cache miss takes around 340 units on the same system).

\end{document}